\documentclass[lettersize,journal]{IEEEtran}

\usepackage{tabularx}
\usepackage{array}
\usepackage{cite}
\usepackage{orcidlink}
\usepackage{amsmath,amssymb,amsfonts}
\usepackage{algorithmic}
\usepackage{graphicx}
\usepackage{textcomp}
\usepackage{subcaption}
\usepackage{xcolor}
\usepackage{amsmath}
\usepackage{algorithm}
\usepackage{algorithmic}
\usepackage{acronym}
\usepackage[inline]{enumitem}

\newcommand{\sectionname}{Sect.}
\newcommand{\meter}{\,\mathrm{m}}
\newcommand{\mW}{\,\mathrm{mW}}
\newcommand{\dB}{\,\mathrm{dB}}
\newcommand{\MHz}{\,\mathrm{MHz}}

\newcommand{\ms}{\,\mathrm{ms}}
\newcommand{\second}{\,\mathrm{s}}

\acrodef{AWGN}{additive white Gaussian noise}
\acrodef{AdamW}{adaptive moment estimation with weight decay}
\acrodef{B5G}{beyond-5G}
\acrodef{CSA}{central satellite}
\acrodef{CFmMIMO}{cell-free massive \acs{MIMO}}
\acrodef{mMIMO}{massive \acs{MIMO}}
\acrodef{DPD} {disk Poisson Process} 
\acrodef{DRA}{direct radiating antenna}
\acrodef{DC}{direct current}
\acrodef{DoD}{depth of discharge}
\acrodef{eMBB}{enhanced mobile broadband}
\acrodef{E-GPW}{extended gridded population of world database}
\acrodef{FDM}{frequency division multiplexing}
\acrodef{FIR}{finite impulse response}
\acrodef{TFoA}{thinned \acs{FoA}}
\acrodef{FoA}{formation of arrays}
\acrodef{GEO}{geostationary Earth orbit}
\acrodef{GSO}{geosynchronous Earth orbit}
\acrodef{H-RRM}{heuristic radio resource management}
\acrodef{KPI}{key performance indicator}
\acrodef{IF}{intermediate frequency}
\acrodef{LEO}{low Earth orbit}
\acrodef{LoS}{line-of-sight}
\acrodef{MAI}{multiple access interference}
\acrodef{MB}{multi beam}
\acrodef{MEO}{medium Earth orbit}
\acrodef{mMTC}{massive machine-type communications}
\acrodef{eMTC}{enhanced machine-type communications}
\acrodef{PDD}{Poisson disk distribution}
\acrodef{EIRP}{effective isotropic radiated power}
\acrodef{NTN}{non terrestrial network}
\acrodef{URLLC}{ultra-reliable low-latency communications}
\acrodef{ECEF}{Earth-centered Earth-fixed}
\acrodef{GPS}{global positioning system}
\acrodef{HTFS}{high throughput fractionated satellite}
\acrodef{UHF}{ultra-high frequency}
\acrodef{FF}{formation flying}
\acrodef{MIMO}{multiple input multiple output}
\acrodef{M-MIMO}{massive multiple input multiple output}
\acrodef{GNC}{guidance navigation and control}
\acrodef{GNSS}{global navigation satellite system}
\acrodef{OFDM}{orthogonal frequency division multiplexing}
\acrodef{PHY}{physical-layer}
\acrodef{SA}{sub-array}
\acrodef{SADM}{solar array drive mechanism}
\acrodef{SG}{solar generator}
\acrodef{SNR}{signal-to-noise ratio}
\acrodef{SIR}{signal-to-interference ratio}
\acrodef{SINR}{signal-to-interference-plus-noise ratio}
\acrodef{SSPA}{solid-state power amplifier}
\acrodef{RF}{radio frequency}
\acrodef{R-GEO}{regional \acs{GEO}}
\acrodef{UT}{user terminal}
\acrodef{UE}{user equipment}
\acrodef{QVD}{quantized virtual distancing}
\acrodef{TDM}{time division multiplexing}
\acrodef{R-GEO}{regional GEO}
\acrodef{RRM}{radio resource management}
\acrodef{SSPA}{solid-state power amplifier}
\acrodef{BLER}{block error rate}
\acrodef{DL}{downlink}
\acrodef{UL}{uplink}
\acrodef{PSD}{power spectral density}
\acrodef{UDSM}{ultra deep sub-micron}
\acrodef{PEPS}{power energy platform simulation}
\acrodef{SE}{spectral efficiency}
\acrodef{wrt}{with respect to}
\acrodef{OBP}{on-board digital processor}
\acrodef{UDSM}{ultra-deep-sub-micron}

\acrodef{3GPP}{third generation partnership project}
\acrodef{AWGN}{additive white Gaussian noise}
\acrodef{B5G}{beyond-5G}
\acrodef{CS}{central satellite}
\acrodef{DRA}{direct radiating antenna}
\acrodef{DC}{direct current}
\acrodef{DoD}{depth of discharge}
\acrodef{eMBB}{enhanced mobile broadband}
\acrodef{FDM}{frequency division multiplexing}
\acrodef{FIR}{finite impulse response}
\acrodef{TFoA}{thinned \acs{FoA}}
\acrodef{FoA}{formation of arrays}
\acrodef{GEO}{geostationary Earth orbit}
\acrodef{GSO}{geosynchronous Earth orbit}
\acrodef{KPI}{key performance indicator}
\acrodef{IF}{intermediate frequency}
\acrodef{LEO}{low Earth orbit}
\acrodef{LoS}{line-of-sight}
\acrodef{MAI}{multiple access interference}
\acrodef{MEO}{medium Earth orbit}
\acrodef{mMTC}{massive machine-type communications}
\acrodef{eMTC}{enhanced machine-type communications}
\acrodef{EIRP}{effective isotropic radiated power}
\acrodef{NTN}{non terrestrial network}
\acrodef{URLLC}{ultra-reliable low-latency communications}
\acrodef{ECEF}{Earth-centered Earth-fixed}
\acrodef{GPS}{global positioning system}
\acrodef{HTFS}{high throughput fractionated satellite}
\acrodef{HTS}{high throughput satellite}
\acrodef{UHF}{ultra-high frequency}
\acrodef{FF}{formation flying}
\acrodef{MIMO}{multiple input multiple output}
\acrodef{GNC}{guidance navigation and control}
\acrodef{NTN}{non-terrestrial network}
\acrodef{GNSS}{global navigation satellite system}
\acrodef{OFDM}{orthogonal frequency division multiplexing}
\acrodef{PHY}{physical-layer}
\acrodef{SA}{satellite array}
\acrodef{SADM}{solar array drive mechanism}
\acrodef{SG}{solar generator}
\acrodef{SNR}{signal-to-noise ratio}
\acrodef{SIR}{signal-to-interference ratio}
\acrodef{SINR}{signal-to-interference-plus-noise ratio}
\acrodef{SSPA}{solid-state power amplifier}
\acrodef{RF}{radio frequency}
\acrodef{R-GEO}{regional \acs{GEO}}
\acrodef{UT}{user terminal}
\acrodef{UE}{user equipment}
\acrodef{TDM}{time division multiplexing}
\acrodef{RRM}{radio resource management}
\acrodef{SSPA}{solid-state power amplifier}
\acrodef{BLER}{block error rate}
\acrodef{DL}{down-link}
\acrodef{UL}{up-link}
\acrodef{RV}{random variable}
\acrodef{PSD}{power spectral density}
\acrodef{UDSM}{ultra deep sub-micron}
\acrodef{PEPS}{power energy platform simulation}
\acrodef{SE}{spectral efficiency}
\acrodef{wrt}{with respect to}
\acrodef{OBP}{on-board digital processor}
\acrodef{UDSM}{ultra-deep-sub-micron}
\acrodef{UC}{user-centric}
\acrodef{CSI}{channel state information}
\acrodef{PAC}{per-antenna constraint}
\acrodef{FPAC}{fair \acs{PAC}}
\acrodef{MPC}{maximum power constraint}
\acrodef{AWGN}{additive white Gaussian noise}
\acrodef{TX}{transmit}
\acrodef{MMSE}{minimum mean square error}
\acrodef{MF}{matched filter}
\acrodef{ZF}{zero forcing}
\acrodef{ELSA}{enhanced logarithmic spiral array}
\acrodef{UPA}{uniform planar array}
\acrodef{NB}{narrowband}
\acrodef{WB}{wideband}
\acrodef{BFN}{beamforming network}
\acrodef{MD-MIQP}{minimum distance mixed integer quadratic problem}
\acrodef{PDF}{probability density function}
\acrodef{NPR}{noise-to-power ratio}
\acrodef{HPA}{high power amplifier}
\acrodef{CF}{cell-free}
\acrodef{3GPP}{third generation partnership project}
\acrodef{UC-MIMO}{user centric MIMO}
\acrodef{VHTS}{very high throughput satellites}
\acrodef{PM-MIMO}{pragmatic M-MIMO}
\acrodef{NR}{new radio}
\acrodef{LMS}{land mobile satellite}
\acrodef{IM}{intermodulation}
\acrodef{OBO}{output back-off}
\acrodef{FDD}{frequency division duplexing}
\acrodef{TDD}{time division duplexing}
\acrodef{BLER}{block error rate}
\acrodef{rv}{random variable}
\acrodef{COB}{center of beam}
\acrodef{ISL}{inter-satellite link}
\acrodef{AP}{access point}
\acrodef{SVD}{single value decomposition}
\acrodef{CW}{continuous wave}
\acrodef{AOCS}{attitude and orbit control system}
\acrodef{LSTM}{long short-term memory}
\acrodef{CPU}{central processing unit}
\acrodef{GPU}{graphics processing unit}
\acrodef{TPU}{tensor processing unit}
\acrodef{MMF}{max-min fairness}
\acrodef{MHA}{multi-head attention}
\acrodef{FFN}{feed-forward network}
\acrodef{MSE}{mean square error}
\acrodef{CDF}{cumulative distribution function}
\acrodef{EPA}{equal power allocation}
\acrodef{FPA}{fractional power allocation}
\acrodef{TNN}{transformer neural network}
\acrodef{RT}{real-time}

\begin{document}

\title{Transformer-Based Power Optimization for  Max-Min Fairness in Cell-Free Massive MIMO}
	\author{Irched Chafaa\,\orcidlink{0000-0003-1467-5933}, 
            Giacomo Bacci\,\orcidlink{0000-0003-1762-8024},~\IEEEmembership{Senior Member,~IEEE},
		Luca Sanguinetti\,\orcidlink{0000-0002-2577-4091},~\IEEEmembership{Fellow,~IEEE}\vspace*{-0.5cm}
		\thanks{This work was supported by the Italian Ministry of Education and Research (MUR) in the framework of the FoReLab Project (Department of Excellence) and in part by the European Union under the Italian National Recovery and Resilience Plan (NRRP) of NextGenerationEU, partnership on ``Telecommunications of the Future'' (PE00000001 -- Program ``RESTART'', Structural Project 6GWINET, Cascade Call SPARKS).}%
		\thanks{I. Chafaa, G. Bacci, L. Sanguinetti are with the Dipartimento di Ingegneria dell'Informazione, University of Pisa, 56122 Pisa, Italy (e-mail: irched.chafaa@ing.unipi.it, \{giacomo.bacci, luca.sanguinetti\}@unipi.it).}
	}



\maketitle

\begin{abstract}
Power allocation is an important task in wireless communication networks. Classical optimization algorithms and deep learning methods, while effective in small and static scenarios, become either computationally demanding or unsuitable for large and dynamic networks with varying user loads. This letter explores the potential of transformer-based deep learning models to address these challenges. We propose a \acl{TNN} to jointly predict optimal uplink and downlink power using only user and access point positions. The max-min fairness problem in cell-free massive MIMO systems is considered. Numerical results show that the trained model provides near-optimal performance and adapts to varying number of users and access points without retraining, additional processing, or updating its neural network architecture. This demonstrates the effectiveness of the proposed model in achieving robust and flexible power allocation for dynamic networks. 
\end{abstract}


\acresetall 
\vspace*{-0.6cm}
\section{Introduction} 
\IEEEPARstart{P}{ower} allocation is a crucial step in wireless networks to optimize the communications performance \cite{tan2017resource}. Depending on the nature of the optimization problem, power allocation can be performed via iterative methods \cite{farooq2020accelerated} such as in the sum \ac{SE} problem \cite{demir2021foundations} or using closed-form solution, as reported in \cite{miretti2022closed} for the particular \ac{MMF} problem. Iterative optimization algorithms require multiple iterations to find optimal powers, often failing to converge within channel coherence time for highly dynamic networks. They also impose high computational complexity due to numerous parameters that scale poorly with network size. In addition, if the configuration changes, as commonly happens in wireless networks, additional iterations are required. The closed-form solution \cite{miretti2022closed} still requires substantial \ac{RT} channel information and is computationally expensive, due to matrix inversion and eigenvalue decomposition. Thus, while these methods provide theoretical benchmarks, they struggle with \ac{RT}, dynamic, and large-scale networks.
   
Machine learning, including reinforcement and deep learning, has been explored to overcome iterative solution limitations \cite{kim2023survey,mao2018deep,kocharlakota2024pilot}. While effective for fixed configurations and low-dimensional inputs, these methods are not suited to varying input sizes and dynamic systems, in the absence of retraining and architectural adjustments. In \cite{kocharlakota2024pilot}, the authors used a transformer-based \cite{vaswani2017attention} model for down-link (DL)  power allocation that adapts to varying numbers of \acp{UE} via unsupervised learning. However, the proposed method requires post-processing and padding, which increases computational overhead and limits scalability, especially with frequent changes in user load \cite{dwarampudi2019effects,alrasheedi2023padding}. In addition, it does not address varying numbers of \acp{AP}. This raises a key question: \emph{How can we design a flexible learning model that can handle different \ac{UE} loads and active number of \acp{AP} while maintaining near-optimal performance?}

In this letter, we propose a supervised learning approach to train a \ac{TNN} \cite{vaswani2017attention} that leverages \ac{UE} and \ac{AP} location information (e.g., spatial coordinates) to jointly predict \ac{UL} and DL powers, maximizing the minimum \ac{SE}. We focus on the \ac{MMF} problem in cell-free \ac{mMIMO} systems \cite{demir2021foundations}, a key candidate for future 6G networks. Our main contributions are as follows. By leveraging the attention mechanism of the transformer, our model efficiently captures \ac{UE}-\ac{AP} relationships in parallel and adapts to the dynamic configuration of the network. Unlike previous works (e.g., \cite{mao2018deep, kocharlakota2024pilot, miretti2022closed}), we use only \ac{UE} and \ac{AP} coordinates (thus neglecting large-scale fading and channel statistics) while maintaining near-optimal performance. This minimal input reduces overhead, enables \ac{RT} operation, and decouples power control from data detection, enhancing its applicability across different network layers. Our approach further addresses a key limitation of classical and learning-based power allocation methods, which require redesigning and retraining for variations in the number of \acp{UE} ($K$) and \acp{AP} ($L$). By adapting the transformer with dynamic input and output layers, our solution seamlessly handles varying $K$ and $L$ without architectural changes, retraining, or additional processing such as padding. Trained on diverse configurations of $K$ and $L$, the model generalizes effectively to unseen setups during inference. Finally, we validate the performance against the optimal closed-form max-min \ac{SE} solution \cite{miretti2022closed}, achieving comparable results with three distinct advantages: increased flexibility, reduced computational complexity, and reduced input information.
 \vspace*{-0.2cm} 
\section{System Model and Problem Formulation}\label{sec:systemModel}
We consider a cell-free \ac{mMIMO} system, where $K$ single-antenna \acp{UE} are served by $L$ \acp{AP} with $N$ antennas each. The \acp{AP} coordinate via a fronthaul network and a \ac{CPU} for joint processing and power allocation. The standard \ac{TDD} protocol of cell-free \ac{mMIMO} is used \cite{demir2021foundations}, where the $\tau_c$
available channel uses are employed for: \ac{UL} training phase ($\tau_p$); \ac{DL} payload transmission ($\tau_d$); and \ac{UL} payload transmission ($\tau_u$). Clearly, $\tau_c\geq\tau_p + \tau_d + \tau_u$. We consider a narrowband channel model and assume that the channel remains constant within a coherence block. We denote the channel vector between the \ac{AP} $l$ and \ac{UE} $k$ with $\mathbf{h}_{lk}$, and model it as \cite{demir2021foundations}:
\begin{align}\label{eq:channel_model}
  \mathbf{h}_{lk} = \sqrt{\beta_{lk}} \mathbf{R}_{lk}^{1/2} \mathbf{g}_{lk}
\end{align}
where $\beta_{lk}$ is the large-scale fading coefficient, accounting for path loss and shadowing, $\mathbf{R}_{lk} \in \mathbb{C}^{N \times N}$ is the spatial correlation matrix at \ac{AP} $l$, and $\mathbf{g}_{lk} \sim \mathcal{CN}(\mathbf{0}, \mathbf{I}_N)$ is an i.i.d. complex Gaussian vector representing the small-scale fading. We assume that the channels $\{\mathbf{h}_{lk}; l=1,\ldots,L\}$ are independent and call $\mathbf{h}_{k} = \left[\mathbf{h}_{1k}^T, \ldots, \mathbf{h}_{Lk}^T \right]^T \in \mathbb{C}^{LN}$ the collective channel from all \acp{AP} to \ac{UE} $k$.

The \ac{CPU} computes the estimate of $\mathbf{h}_{k}$ on the basis of received pilot sequences transmitted during the training phase \cite{demir2021foundations}. The \ac{MMSE} estimate is $\widehat{\mathbf{h}}_{k} = [\widehat{\mathbf{h}}_{1k}^T, \ldots, \widehat{\mathbf{h}}_{Lk}^T ]^T$ with \cite{demir2021foundations}
\begin{align}
\widehat{\mathbf{h}}_{lk} = \mathbf{R}_{lk} \mathbf{Q}_{lk}^{-1} \left( \mathbf{h}_{l k} +  \frac{1}{\tau_p \rho} \mathbf{n}_{lk} \right) \sim \mathcal{N}_C \left( \mathbf{0}_N, \mathbf{\Phi}_{lk} \right)
\end{align}
where $\rho$ is the \ac{UL} pilot power of each \ac{UE}, $\mathbf{n}_{lk} \sim \mathcal{N}_C(\mathbf{0}_{LN}, \sigma^2\mathbf{I}_{LN})$ is the thermal noise, and $\mathbf{\Phi}_{lk} = \mathbf{R}_{lk} \mathbf{Q}_{lk}^{-1} \mathbf{R}_{lk}$, where $\mathbf{Q}_{lk} = \mathbf{R}_{l k} + \frac{\sigma^2}{\tau_p\rho} \mathbf{I}_{LN}$. Hence, $\widehat{\mathbf{h}}_{k} \sim \mathcal{N}_C \left( \mathbf{0}_{LN}, \mathbf{\Phi}_{k} \right)$, with $\mathbf{\Phi}_{k} = {\mathrm{diag}}(\mathbf{\Phi}_{1k},\ldots,\mathbf{\Phi}_{Lk})$. Note that the method proposed in this letter can be applied to other channel estimation schemes, such as the least-squares one \cite{demir2021foundations}. 
\vspace*{-0.2cm} 
\subsection{Uplink and Downlink Transmissions}
To detect the data of \ac{UE} $k$ in the \ac{UL}, the \ac{CPU} selects an arbitrary receive combining vector $\mathbf{v}_{k}\in \mathbb{C}^{LN}$ based on all the collective channel estimates $\{\widehat{\mathbf{h}}_{k}; k =1,\ldots, K\}$. An achievable \ac{SE} of \ac{UE} $k$ is given by \cite{demir2021foundations}: 
\begin{align}\label{eq:spectral_efficiency_uplink}
  \text{\ac{SE}}_k^\text{UL} = \frac{\tau_u}{\tau_c} \log_2(1 + \text{SINR}_k^\text{UL})
\end{align}
with the effective \ac{SINR} defined as
 \begin{align}
\!\!\!\!\!\frac{p_k^\text{UL} \left| \mathbb{E} \left\{ \mathbf{v}_{k}^\mathrm{H} \mathbf{h}_{k} \right\} \right|^2}{\sum\limits_{i=1}^K p_i^\text{UL} \mathbb{E} \left\{ \left| \mathbf{v}_{k}^\mathrm{H} \mathbf{h}_{i} \right|^2 \right\} - p_k^\text{UL} \left| \mathbb{E} \left\{ \mathbf{v}_{k}^\mathrm{H} \mathbf{h}_{k} \right\} \right|^2 + \sigma^2\mathbb{E} \left\{ \| \mathbf{v}_k \|^2 \right\}}
\end{align}
where $p_k^\text{UL}$ is the \ac{UL} transmit power of user $k$. The expectation is taken with respect to all sources of randomness. Although the bound in \eqref{eq:spectral_efficiency_uplink} is valid for any combining vector, we consider the \ac{MMSE} combiner, given by\cite{demir2021foundations}: 
\begin{align}
\mathbf{v}_{k} = \left( \sum_{k=1}^{K} p_k^\text{UL} \widehat{\mathbf{h}}_{k} \widehat{\mathbf{h}}_{k}^H + \mathbf{Z} \right)^{-1} \widehat{\mathbf{h}}_{k} 
\end{align}
where $\mathbf{Z} =  \sum_{k=1}^{K} p_k^\text{UL}(\mathbf{R}_{k} - \mathbf{\Phi}_{k}) + \sigma^2 \mathbf{I}_{LN}$.

In the \ac{DL}, the \ac{CPU} coordinates the \acp{AP} to transmit signals to the UEs. Similarly to \ac{UL}, an achievable \ac{SE} of user $k$ is obtained as:
\begin{align}
  \text{ \ac{SE}}_k^\text{DL} = \frac{\tau_d}{\tau_c}\log_2(1 + \text{SINR}_k^\text{DL})
\end{align}
with the effective \ac{SINR} defined as
 \begin{align}
\frac{p_k^\text{DL} \left| \mathbb{E} \left\{ \mathbf{h}_{k}^\mathrm{H} \mathbf{w}_{k} \right\} \right|^2}{\sum\limits_{i=1}^K p_i^\text{DL} \mathbb{E} \left\{ \left| \mathbf{h}_{k}^\mathrm{H} \mathbf{w}_{i} \right|^2 \right\} - p_k^\text{DL} \left| \mathbb{E} \left\{ \mathbf{h}_{k}^\mathrm{H} \mathbf{w}_{k} \right\} \right|^2 + \sigma^2}
\end{align}
where $p_{k}^\text{DL}$ is the \ac{DL} power used by the \ac{CPU} to serve \ac{UE} $k$ and $\mathbf{w}_{k}\in \mathbb{C}^{LN}$ is its associated unit-norm precoding vector. The \ac{MMSE} precoder is used \cite{demir2021foundations}, which is given by $
\mathbf{w}_{k} = \frac{\mathbf{v}_{k}}{\|\mathbf{v}_{k}\|}$.
\vspace*{-0.2cm} 
\subsection{Problem Formulation}
We aim to develop a flexible, learning-based solution for optimal power allocation in networks with varying $K$ and $L$, using minimal input information. In particular, we consider the max-min optimization problem, commonly used in cell-free \ac{mMIMO} systems to ensure equal \ac{SE} across \acp{UE}. In the \ac{UL}, the problem takes the following form \cite{demir2021foundations}:
\begin{align}
\begin{aligned}
  \max_{\{p_k^\text{UL}\geq 0\}} \min_k &\ \textrm{\ac{SE}}_k^\text{UL} \\
  \text{subject to} &\quad  p_k^\text{UL} \leq P_{k,\max}^\text{UL} \ \forall k
\end{aligned}
\label{optul}
\end{align}
where $P_{k,\max}^\text{UL}$ is the maximum \ac{UL} power for user $k$. Similarly, in the \ac{DL} we have that:
\begin{align}
\begin{aligned}
  \max_{\{p_k^\text{DL}\geq 0\}} \min_k &\ \text{\ac{SE}}_k^\text{DL}\\
  \text{subject to} &\quad \textstyle{\sum_{k=1}^K p_k^\text{DL} \leq \sum_{l=1}^L P_{l,\max}^\text{DL}}
\end{aligned}
\label{optdl}
\end{align}
where $P_{l,\max}^\text{DL}$ is the maximum power per \ac{AP}. The constraint ensures that the total power allocated to all \acp{UE} does not exceed the total power budget across all \acp{AP}.

Both optimization problems can be solved using the closed-form solution in \cite{miretti2022closed}, online iterative solvers \cite{farooq2020accelerated,demir2021foundations}, or traditional \ac{DL} models \cite{kim2023survey,mao2018deep,kocharlakota2024pilot}. However, as discussed earlier, all these methods may be demanding for \ac{RT} application in dynamic situations. To address this, we propose a flexible, data-driven alternative: a supervised learning framework with a trained \ac{TNN} that jointly predicts the optimal powers, while handling varying numbers of \acp{UE} and \acp{AP}.
\section{Transformer-based Power Allocation}
Unlike other deep learning architectures \cite{kim2023survey,mao2018deep,kocharlakota2024pilot}, transformers can handle size-varying inputs and outputs without the need for additional processing steps \cite{vaswani2017attention}. The parallel processing capability of transformers further enhances their efficiency, making them ideal for \ac{RT} applications. In addition, the self-attention mechanism  enables them to capture complex relationships between UEs, \acp{AP}, and channels, enabling more accurate and scalable power allocation predictions. By training on data spanning different numbers of \ac{UE} and \ac{AP} configurations, transformers can generalize well across a wide range of scenarios, eliminating the need to retrain or reconfigure the model's architecture for each new scenario. 
In the following, we explain in detail the basic parts of our proposed solution.

\subsection{Training Data}

For a given \ac{AP} placement and path loss model, we generate multiple \ac{UE} configurations to evaluate large-scale fading coefficients. Optimal \ac{UL} and \ac{DL} powers are then computed offline by solving the max-min \ac{SE} optimization problems in \eqref{optul} and \eqref{optdl}, using  the closed-form solution in \cite{miretti2022closed}. The training dataset consists of input-output pairs $(\bf{Z}, \mathbf{p}^{\star})$, where ${\bf Z} \in \mathbb{R}^{K \times (2L+2)}$ captures normalized $x$-$y$ coordinates of $K$ \acp{UE} and $L$ \acp{AP}, and $\mathbf{p}^{\star} \in \mathbb{R}^{2K}$ contains the \ac{UL} and \ac{DL} optimal power values. Min-max normalization \cite{mao2018deep} ensures consistent scaling for stable training. The dataset is split $80$-$20$ into training and testing sets. To enhance generalization, the dataset covers diverse configurations, varying \ac{UE} and \ac{AP} counts, distributions, and channel realizations. Multiple samples per $(K, L)$ combination enable learning robust power allocation across cell-free \ac{mMIMO} scenarios.
\subsection{Model Architecture}

\begin{figure}[t]
  \centering
 \includegraphics[width=0.95\columnwidth]{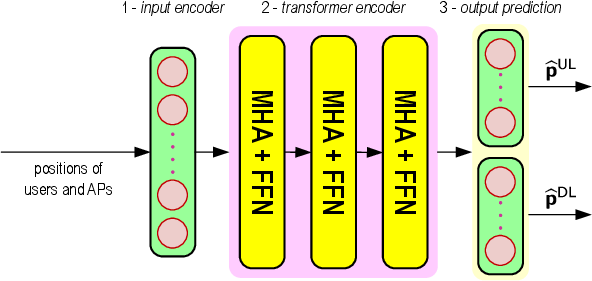}
  \caption{Architecture diagram of the proposed transformer-based model to predict jointly \ac{UL} and \ac{DL} powers leveraging spatial information at the input.}
  \label{arch}
  \end{figure}

The proposed \ac{TNN} model predicts \ac{UL} and \ac{DL} powers by capturing relationships between input features. It comprises three key components: a \emph{dynamic} input layer, a multi-layer transformer encoder, and an output layer, enabling adaptability to varying \acp{UE} and \acp{AP}. \figurename~\ref{arch} illustrates the processing steps.

\begin{enumerate}
\item \emph{Dynamic input layer}: For each batch of size $B$, the input tensor $ \mathbf{X} \in \mathbb{R}^{B \times K \times (2L+2)}$ encodes spatial information for $K$ \acp{UE} and $L$ \acp{AP}, including only their $x$-$y$ coordinates. A fully connected input layer maps features to a $ d_{\text{mod}}$-dimensional space using \cite{vaswani2017attention,lin2022survey}: 
\begin{align}
\mathbf{H} = \text{ReLU}(\mathbf{X} \mathbf{W}_\text{input} + \mathbf{b}_\text{input})
\end{align}
where $\text{ReLU}$ is an activation function \cite{rasamoelina2020review}, $ \mathbf{W}_\text{input} \in \mathbb{R}^{(2L+2) \times d_{\text{mod}}} $ and $ \mathbf{b}_\text{input} \in \mathbb{R}^{d_{\text{mod}}}$ are learnable parameters. The resulting tensor $ \mathbf{H} \in \mathbb{R}^{B \times K \times d_{\text{mod}}} $ is fed into the transformer encoder. Unlike previous methods \cite{kim2023survey,mao2018deep,kocharlakota2024pilot}, our model dynamically adjusts the input tensor size $\mathbf{X}$ based on $K$ and $L$ in each batch. This enables seamless handling of varying $K$ and $L$ during both training and inference, eliminating the need to add more processing steps. Note that, while the model does not explicitly receive channel information as an input, its effect is accounted for indirectly through the training labels $\mathbf{p}^{\star}$, which are computed offline using a channel-aware solution for each \ac{UE}/\ac{AP} position.
\item \emph{Multi-layer transformer encoder}: The core of the model is a transformer encoder that processes the tensor $\mathbf{H}$. It comprises $M$ layers, each featuring a \ac{MHA} mechanism and a \ac{FFN} \cite{vaswani2017attention}. The self-attention mechanism effectively captures dependencies by computing attention scores between all input element pairs:
\begin{align}
  \text{Attention}(\mathbf{Q}, \mathbf{K}, \mathbf{V}) = \text{Softmax}\left(\frac{\mathbf{Q}\mathbf{K}^\top}{\sqrt{D_K}}\right)\mathbf{V}
\end{align}
where $\text{Softmax}$ is an activation function \cite{rasamoelina2020review}, $\mathbf{Q}$, $\mathbf{K}$, and $\mathbf{V}$ are query, key, and value matrices derived from $\mathbf{H}$ with $D_K$ being a dimension of the key matrix \cite{vaswani2017attention,lin2022survey}. After that, each \ac{UE}'s feature is processed through the \ac{FFN}:
\begin{align}
  \text{FFN}(\mathbf{h}) = \text{ReLU}(\mathbf{h}\mathbf{W}_1 + \mathbf{b}_1)\mathbf{W}_2 + \mathbf{b}_2
\end{align}
where $\mathbf{W}_1, \mathbf{W}_2, \mathbf{b}_1, \mathbf{b}_2$ are trainable parameters. As a result, the transformer encoder output, $\mathbf{H}_\text{out} \in \mathbb{R}^{B \times K \times d_{\text{mod}}}$, encodes the learned  relationships between all \acp{UE} and \acp{AP}, effectively capturing their interactions.

\item \emph{Output layer}:
To predict the \ac{UL} and \ac{DL} powers, the output of the transformer encoder $\mathbf{H}_\text{out}$ is passed through two separate fully-connected layers yielding:
\begin{align}
\hat{p}_k^{\text{UL}} &= \text{Sigmoid}\left(\mathbf{H}_{\text{out}} \mathbf{W}_{\text{UL}} + \mathbf{b}_{\text{UL}} \right) \cdot P_{k,\max}^\text{UL}\\
\hat{p}_k^{\text{DL}} &= \text{ReLU}\left(H_{\text{out}} \mathbf{W}_{\text{DL}} + \mathbf{b}_{\text{DL}}\right) \cdot \frac{\sum_{l=1}^L P_{l,\max}^\text{DL}}{\sum_{k=1}^K \hat{p}_k^\text{DL}}
\end{align}
where $\mathbf{W}_{\text{UL}}, \mathbf{W}_{\text{DL}}, \mathbf{b}_{\text{UL}}, \mathbf{b}_{\text{DL}}$ are trainable parameters, and $\text{Sigmoid}$ is an activation function ensuring an output in the range $[0,1]$ \cite{rasamoelina2020review}. The predicted \ac{UL} and \ac{DL} powers for each user are concatenated to form the final output tensor of powers $\widehat{\mathbf{p}} \in \mathbb{R}^{B \times K \times 2}$ for all \acp{UE} in the batch. 
\end{enumerate}

The model is trained to minimize the \ac{MSE} between the predicted powers $\widehat{\mathbf{p}}$ and the optimal powers $\mathbf{p}^{\star}$ from offline optimization. By doing so, the model implicitly encodes the effects of the channel propagation environment during training, and learns to approximate the function, mapping the positions of \acp{UE} and \acp{AP} to optimal powers that maximize the minimum \ac{SE}:
\begin{align}
  \mathcal{L}_\text{MSE} = \frac{1}{B} \sum_{i=1}^B \|\mathbf{p}_i^{\star} - \widehat{\mathbf{p}}_i\|^2.  
\end{align}
\vspace*{-0.4cm} 
\subsection{Computational Complexity}
The computational complexity of the proposed  model differs between the training and inference phases. During training, the model performs both a forward and backward pass, with the main cost arising from the self-attention mechanism in the transformer encoder. The overall training complexity  \cite{vaswani2017attention,lin2022survey} is in the order of:
\begin{align}
\mathcal{O}\left( 2 M B  K d_{\text{mod}} \left(  d_{\text{mod}} +  K \right)+ 2  B  K  \left( 2 L + 2 \right) d_{\text{mod}} \right).
\end{align} 
In contrast, inference involves only a forward pass, which is faster and more efficient since it lacks back-propagation and gradient updates and only uses $x$-$y$ coordinates as input. The inference complexity per sample \cite{vaswani2017attention,lin2022survey} is in the order of:
\begin{align}
\mathcal{O}\left( M \left(   K d_{\text{mod}}^2 +  K^2 d_{\text{mod}} \right)+   K  \left( 2L + 2 \right) d_{\text{mod}} \right).
\end{align} 

For high numbers of \acp{UE} ($K$), the complexity for predicting UL and DL powers jointly is dominated by $\mathcal{O}(M  d_{\text{mod}} K^2)$. In contrast, the closed-form solution in \cite{miretti2022closed} has a higher complexity of $\mathcal{O}(K^3)$ for each power prediction (UL or DL separately). Iterative optimization methods such as \cite{farooq2020accelerated}, scales with $\mathcal{O}(Q L K^2)$, where $Q$ represents the number of iterations required to converge, which can exceed hundreds \cite{kocharlakota2024pilot}. Moreover, the optimal powers  are computed based on a substantial amount of information, including channel statistics (such as covariance matrices, channel estimates, and large-scale fading) as well as combining and precoding vectors. These elements must be updated regularly according to the current network configuration and \ac{UE} load. As a result, the optimal powers must be recalculated each time the parameters $K$ or $L$ change. All these points establish the computational advantage of our method compared to existing approaches.

To further demonstrate the efficiency of the proposed model, we compare the computational time required to predict both \ac{UL} and \ac{DL} powers with $K=40$ and $L=16$: using the same computational facilities, the proposed method has a runtime of $18.5\ms$, around $1700$ times faster than the one required by the closed-form solution ($31.6\second$).
 \vspace*{-0.2cm} 
\section{Numerical Results}
In this section, we present numerical results to illustrate the performance of the proposed solution for a cell-free \ac{mMIMO} system, as described in \sectionname~\ref{sec:systemModel}.
 
\subsection{Cell-free \ac{mMIMO} parameters}
We consider a network with a coverage area of $500\meter \times 500\meter$, with $N = 4$ antennas per \ac{AP}. The \acp{AP} are uniformly deployed within the squared coverage area. The maximum \ac{UL} transmit power for each user is $100\mW$, whereas the maximum \ac{DL} transmit power for each \ac{AP} is $200\mW$. We assume $\tau_c = 200$ and set $\tau_p = K$, $\tau_u = \left\lfloor \left(\tau_c - \tau_p\right)/2 \right\rfloor $  and $\tau_d = \tau_c -\tau_p - \tau_u$. Large-scale fading coefficients are computed following the 3GPP path-loss model adopted in \cite[\sectionname~III-D]{miretti2022closed} for a $2$-GHz carrier frequency, a pathloss exponent of $3.67$, a \ac{UE}-\ac{AP} height difference of $10\meter$ and a shadow fading $F_{kl} \sim \mathcal{CN}(0, \alpha^2)$, with $ \alpha^2 = 4\dB$. The shadow fading terms are spatially correlated as in \cite[\sectionname~III-D]{miretti2022closed} to account for the fact that closely located \acp{UE} experience similar shadow fading effects. The noise power is $\sigma^2 = -94$ dBm \cite{miretti2022closed} with a noise figure $\eta = 7\dB$ and a bandwidth $B = 20\MHz$.

A total number of $800$ training samples are generated for each value of $K \in \{2,4,6,8,10\}$ and $ L \in \{9,16\} $. Each sample consists of \ac{UE} and \ac{AP} positions, and optimal powers. Additionally, a testing dataset of $200$ samples for  $K= 2, 3, \ldots, 100$ and $L =4, 5, \ldots,49$ is also generated. By extending the testing range beyond the training values for both \acp{UE} and \acp{AP}, the model's ability to generalize to new dynamic network configurations can be effectively assessed.

\subsection{Learning model parameters}
The transformer architecture includes $M=2$ encoder layers with four attention heads and a model dimension $d_{\text{mod}}=32$. The model's training incorporates a dropout rate of $0.1$ to reduce overfitting by randomly deactivating connections during the training. The learning rate is set to $0.001$ for consistent and controlled optimization using the well-known \ac{AdamW} optimizer \cite{zhou2024towards}, which combines fast convergence with improved regularization. The training is performed with $10$ epochs for each chosen value of $K$ and $L$, with a batch size of $32$ samples. 
\begin{figure}[t!]
    \centering
      \begin{subfigure}[b]{0.48\textwidth}
         \centering
\includegraphics[width=\textwidth]{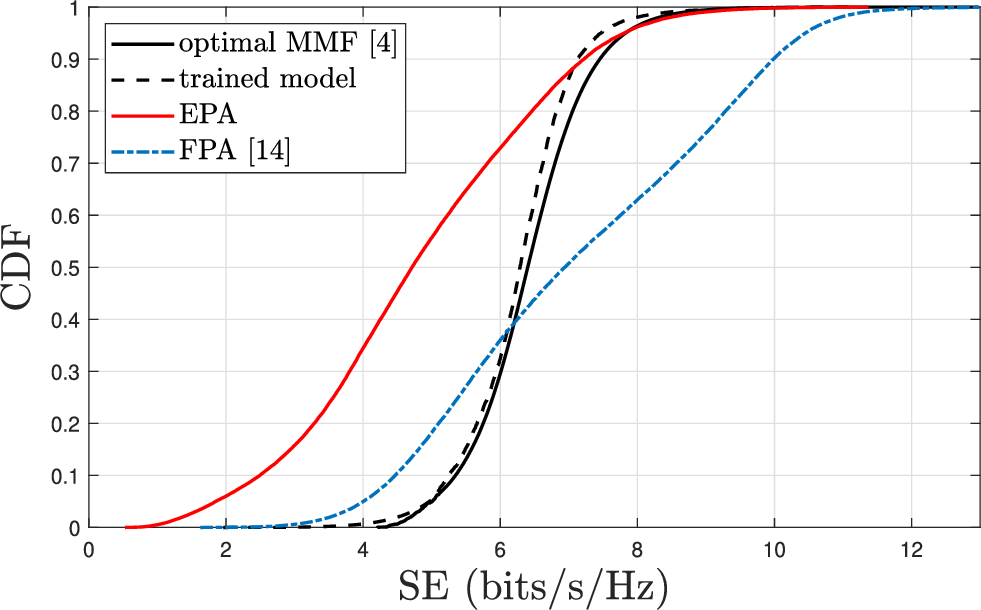}
     \caption{uplink.}
    \label{sim1a}
     \end{subfigure}
     \hfill
     \begin{subfigure}[b]{0.48\textwidth}
         \centering
\includegraphics[width=\textwidth]{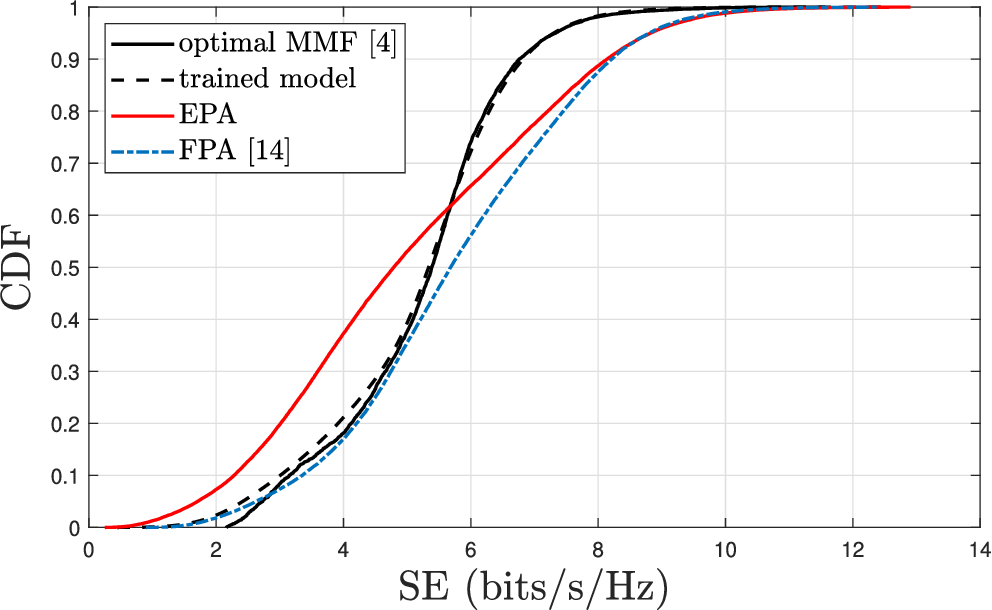}
    \caption{downlink.}
         \label{sim1b}
     \end{subfigure}
    \caption{\acs{CDF} of average per-\acs{UE} \acs{SE} in \acs{UL} and \acs{DL} on the test set of different values of $K$ and $L$. The trained model performs close to the optimal one and generalizes well on different network configurations using the same architecture.}
    \label{fig:NMSE}
\end{figure}
\vspace*{-0.1cm} 
  \subsection{Performance evaluation}
We plot the \ac{CDF} of the average per-\ac{UE} \ac{SE} obtained by evaluating our trained model on test data (unseen during training) across diverse configurations of $2\leq K \leq 100$ and $4 \leq L \leq 36$, including values beyond the training range. The \ac{UL} and \ac{DL} results are shown in \figurename~\ref{sim1a} and \figurename~\ref{sim1b}, respectively, and comparisons are made with  
\begin{enumerate*}[label=\emph{\roman*})]
\item the optimal closed-form solution from \cite{miretti2022closed},
\item \ac{EPA}, and
\item \ac{FPA} \cite{nikbakht2019uplink}.
\end{enumerate*}
The results illustrate that the \ac{SE} achieved by the trained model closely follows the optimal solution, highlighting its ability to deliver near-optimal performance for both \ac{UL} and \ac{DL}. Importantly, the proposed model achieves smaller difference between lower and upper tails of the \ac{CDF}, indicating better fairness among \acp{UE}, unlike \ac{EPA} and \ac{FPA} \acp{CDF} (despite occasional intersections with the optimal \ac{CDF}). These results show not only the ability to predict near-optimal \ac{UL} and \ac{DL} powers, but also a flexibility for different \ac{UE} load and \ac{AP} configurations (using the same model without redesigning its architecture or retraining).

\begin{figure}[t]
  \centering
 \includegraphics[width=\columnwidth]{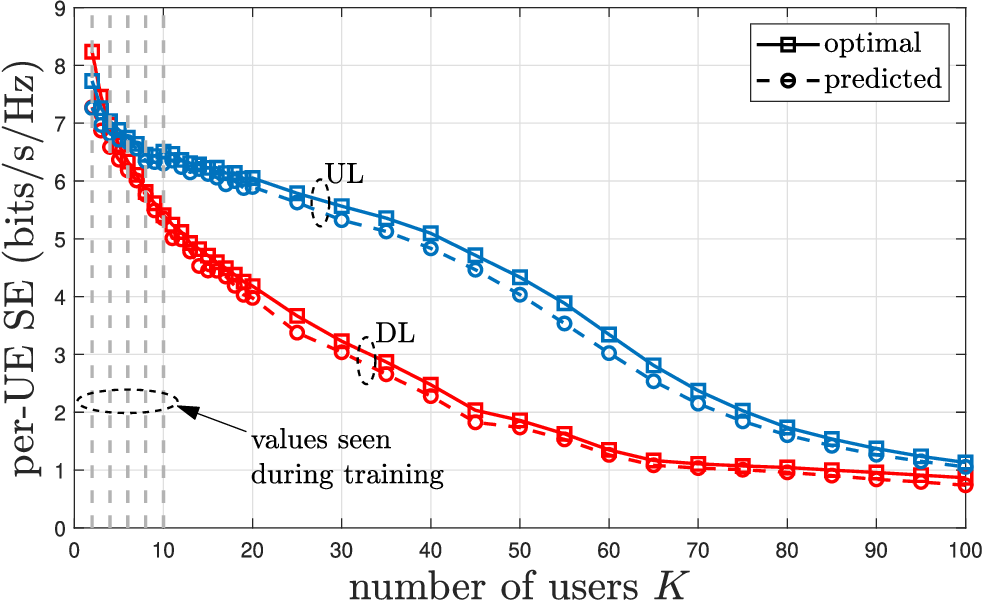}
  \caption{Average per-\acs{UE} \acs{SE} for different $K$ values . The trained model adapts well to  unseen values of $K$ without  retraining or reconfiguring the model’s architecture.}
  \label{sim2}
  \end{figure}

\figurename~\ref{sim2} compares the per-\ac{UE} \ac{SE}, achieved by the optimal and predicted powers, for different values of $K$ and $L=16$ of the test set. First, we notice that the \ac{SE} decreases when $K$ increases, as expected, but gradually, showing that the system handles the increasing user load efficiently. Moreover, \figurename~\ref{sim2} shows that the trained model achieves near-optimal performance even for unseen and higher numbers of \acp{UE} $K>10$. This is achieved using just \ac{UE} and \ac{AP} positions as inputs and without retraining the model, redesigning its architecture, or employing additional data-processing. 

\begin{figure}[t]
  \centering
 \includegraphics[width=\columnwidth]{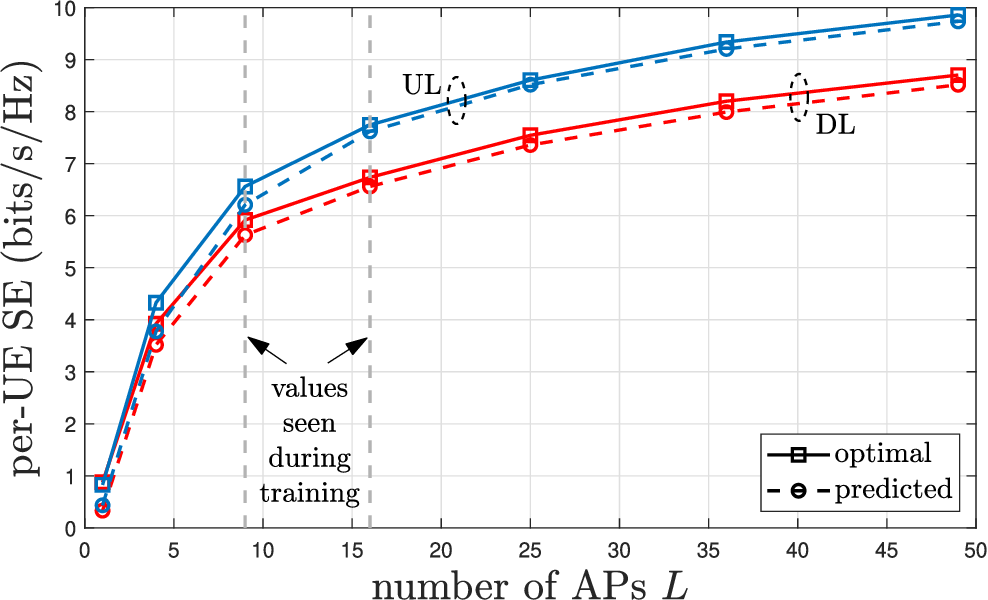}
  \caption{Average per-\acs{UE} \acs{SE} for different numbers of \acsp{AP}. The model handles varying numbers of \acp{AP} without a significant performance degradation.}
  \label{sim3}
  \end{figure}

In \figurename~\ref{sim3}, the trained model is evaluated on the test data with $K=10$ for varying numbers of \acp{AP} $L$ as typically happens in \emph{user-centric} cell-free \ac{mMIMO}. As expected, the per-\ac{UE} \ac{SE} increases as $L$ increases. Furthermore, the \ac{SE} obtained with the trained model follows the optimal one, illustrating a good generalization across different \ac{AP} counts for both \ac{UL} and \ac{DL}. Again, this is achieved without changing the model's architecture or employing additional processing to the data.

 \vspace*{-0.1cm} 
\section{Conclusions}
This letter presented a supervised learning approach to train a \ac{TNN} for jointly predicting \ac{UL} and \ac{DL} powers in a wireless  network, using only the spatial coordinates of \acp{UE} and \acp{AP}. As a case study, the max-min problem in a cell-free \ac{mMIMO} system was considered. The proposed model efficiently handles varying numbers of \acp{UE} and \acp{AP} without requiring retraining or adjustments to the neural network, leveraging the transformer's architecture and dynamic training on diverse configurations. Additionally, the model is designed to address both \ac{UL} and \ac{DL} power allocation through dedicated output layers, which leads to more flexibility. The model is trained, using labels that inherently reflects channel conditions, to learn a mapping that mimics the behavior of the closed-form solution and provide fairness among users. Numerical results showed that our model achieves near-optimal performance across varying system parameters, highlighting its flexibility for dynamic power allocation. While our model has a computation advantage compared to current methods, it still grows quadratically with the number of users due to the attention mechanism in transformers. Thus, to reduce complexity for very large systems, future work will explore efficient transformer variants using sparse or local attention ~\cite{lin2022survey}, which can potentially reduce the complexity from $\mathcal{O}(M  K^2 d_{\text{mod}})$ to linear or logarithmic in $K$.

\bibliographystyle{IEEEtran}
\vspace*{-0.3cm}
\bibliography{IEEEabrv, biblio}

\end{document}